\documentclass[a4paper,11pt]{article}
\pdfoutput=1 
\usepackage{jinstpub} 
\usepackage{color}
\usepackage{url}
\usepackage{subcaption}  
\usepackage{array} 
\usepackage[normalem]{ulem}
\newcolumntype{L}[1]{>{\raggedright\let\newline\\\arraybackslash\hspace{0pt}}m{#1}}
\newcolumntype{C}[1]{>{\centering\let\newline\\\arraybackslash\hspace{0pt}}m{#1}}
\newcolumntype{R}[1]{>{\raggedleft\let\newline\\\arraybackslash\hspace{0pt}}m{#1}}

\definecolor{darkgreen}{RGB}{0,128,0}
\definecolor{gray}{RGB}{64,64,64}
\newcommand\removed[1]{} 
\newcommand\added[1]{#1} 

\newcommand\comment[1]{}

\usepackage{lineno}

\title{Sensitivity analysis towards trace-uranium detection with $\gamma$-$\gamma$ coincidence NAA}

\author[]{R.H.M.~Tsang,}
\author[]{O.~Nusair,} 
\author[]{and A.~Piepke}

\affiliation[]{Department of Physics and Astronomy, University of Alabama, Tuscaloosa, AL 35487}
\emailAdd{hmtsang@ua.edu}
\abstract{We present an improved approach for detection of trace amounts of $^{238}$U by means of neutron activation analysis (NAA). 
The analysis enhancement is obtained by utilizing $\gamma$-$\gamma$ coincidence counting.
An empirical method for evaluating the sensitivity gain in the presence of a large source-related background is presented.
A comparison of detection limits is made between two counting schemes; namely, counting single gammas using one HPGe detector versus counting coincident gammas using two HPGe detectors placed face-to-face.
A data-validated radiation transport model, created in GEANT4 for a single HPGe detector, is extended to handle the two-detector setup. 
In this counting scheme, a $^{238}$U-detection sensitivity enhancement of about a factor of 8 is predicted for a sample of Saint-Gobain G3 Sapphire when compared to the simpler singles counting approach.
}
\arxivnumber{2108.07841} 

\begin{document}
\maketitle
\flushbottom


\section{Introduction}
\label{sec:intro}

Neutron activation analysis (NAA) is a powerful trace-element analysis technique, widely used for radioassay in rare event searches. 
$^{40}$K, $^{232}$Th, and $^{238}$U are often significant sources of background in low-energy searches.
The ability to detect them even at extremely low concentrations in experiment components is often a key for background prevention. 
This paper focuses on the detection of $^{238}$U by NAA.
During neutron activation, $^{238}$U nuclei in the sample are transmuted into a shorter-lived species through $^{238}$U(n,$\gamma$)$^{239}$U. 
Shortly afterwards, $^{239}$U (T$_{1/2}$=23.45 m) undergoes $\beta^-$-decay which produces $^{239}$Np. 
At \removed{high enough}\added{sufficiently high} neutron flux the $^{239}$Np activity exceeds that of the parent $^{238}$U, boosting analysis sensitivity.
The mass and concentration of $^{238}$U contained in the sample can be deduced from the $^{239}$Np activity, measured by means of $\beta$-delayed $\gamma$ spectrometry, knowledge of the neutron flux and energy distribution, and the tabulated reaction cross sections.

The study presented here is aimed at improving the trace element analysis of sapphire (Al$_2$O$_3)$, which is of importance to the proposed nEXO double beta decay experiment~\cite{nexo_sens_2018}.
The method is also applicable to other materials. 
Sapphire cannot be analyzed by inductively coupled plasma mass spectromerty (ICP-MS) as it does not dissolve well in mineral acids.
Analysis sensitivities achievable by NAA are limited by the fast neutron-induced side reaction $^{27}$Al(n,$\alpha$)$^{24}$Na (T$_{1/2}$=14.997 h) 
on the sapphire itself and the presence of chemical impurities, 
leading to the creation of relatively large $^{72}$Ga (T$_{1/2}$=14.1 h) and $^{99}$Mo (T$_{1/2}$=65.976 h) side activities. 
The Compton continua created by the $\gamma$-radiation emitted by these parasitic activities mask the decay of $^{239}$Np, 
thus, limiting the achievable analysis sensitivity. Detection limits for $^{232}$Th, 
via the detection of its activation product $^{233}$Pa (T$_{1/2}$=26.975 d), 
do not suffer from these short-lived by-products because of its relatively long half-life.
Thus, waiting for these dominant side activities to decay away enables sub-ppt levels of Th-analysis sensitivity.

The nuclide $^{239}$Np has two useful $\gamma$-cascades: (106-277) keV and (106-228) keV, that can be exploited in $\gamma$-$\gamma$ coincidence counting. 
With $\gamma$-$\gamma$ coincidence spectrometry, interferences can be greatly reduced by requiring that two gamma rays be detected simultaneously by the two gamma ray detectors.
Background from nuclei that emit single $\gamma$'s would not directly trigger \removed{the}\added{both} detectors. Only Compton-scattered photons with just the right energies could create double triggers.
This technique has previously been employed successfully in the analysis of other elements
\cite{Kim:1965,Vobecky:1999,Hatsukawa:2007}.
Obviously, $\gamma$-$\gamma$ coincidence spectrometry is useful only if the nuclide of interest undergoes a gamma cascade. For this reason, $^{233}$Pa is not a good candidate for coincidence counting.

In this paper, we compare two counting schemes for $^{239}$Np, 
one based on single $\gamma$ spectrometry, and 
the other on $\gamma$-$\gamma$ coincidence spectrometry.
However, achieving a net improvement requires that the achievable background reduction outweighs the loss of counting efficiency. 
A Monte Carlo simulation is, thus, needed to evaluate two figures-of-merit (FOM),
sensitivity and discovery potential, in order to quantify the detection limits of the two analysis methods.
This approach allows to properly account for source related background caused by Compton scattering of parasitic photons and the loss in efficiency due to the requirement of two instead of one full absorption Ge-detector events.

\section{Analysis methods}

In this section, we describe the single $\gamma$ and the $\gamma$-$\gamma$ coincidence counting schemes and the relevant analysis methods employed in this study.

\subsection{Activation}
Both methods utilize samples created by neutron activation in a reactor. 
When the sample is in the reactor, the atomic nuclei in the sample are subjected to a high flux of neutrons. 
Some of the nuclei of the target trace element are transmuted into another isotope, typically the one with an additional neutron,
via the radiative capture reaction $^{A}Z (n,\gamma) ^{A+1}Z$.
The activity of each nuclide $j$ created by the neutron activation process can be described by the following equation:
\begin{equation}
\label{eq:A}
A_{0,j}= \frac{m_j}{M_{j}}\cdot N_{A} \cdot \alpha_j \cdot R_j \cdot (1-e^{\frac{-t_{a}}{\tau_j}} ) 
\end{equation}
where 
$A_{0,j}$ is the activity of the activation product immediately after the irradiation,
$m_j$ is the mass of the target trace element in the sample,
$M_{j}$ is the molar mass of the target trace element,
$N_A$ is Avogadro's number,
$\alpha_j$ is the isotopic abundance of the target nuclide,
$t_a$ is the activation time,
$\tau_j$ is the decay mean life time of the activation product, and 
$R_j$ is the production yield of the activation product. 
$R_j$ can be calculated from the integral neutron fluxes and the energy-spectrum-averaged reaction cross-sections as follows:
\begin{equation}
\label{eq:R}
R_j = \phi_{th} \cdot  \langle \sigma_{th} \rangle_{t} + \phi_{epi} \cdot  \langle \sigma_{epi} \rangle_{t} + \phi_{f} \cdot  \langle \sigma_{f} \rangle_{t}
\end{equation}
where $\phi_{th}$, $\phi_{epi}$, and $\phi_{f}$ are, respectively, the thermal, epithermal, and fast neutron fluxes, 
$\langle \sigma_{th} \rangle_{t}$, 
$\langle \sigma_{epi} \rangle_{t}$, and
$\langle \sigma_{f} \rangle_{t}$ are the corresponding 
neutron-induced reaction
cross-sections 
averaged over the
appropriate neutron energy distributions for the thermal, epithermal, and
fast-neutron flux components 
taken e.g. from the JENDL-4.0 nuclear data library \cite{jendl}.
For more details of the activation process, refer to Appendix A of \cite{EXO200:2017}.

\subsection{Counting}
After irradiation and transportation from the reactor to the counting facility, 
the activated sample is counted with one or several $\gamma$-ray detectors.
For the case of single gamma spectrometry,
the sample is counted with a single HPGe detector for a set period of time.
The energy, intensity, and half-life of the gamma ray emissions from the activated sample 
are measured, and then used to identify the nuclide and quantify its activity.
For $\gamma$-$\gamma$ coincidence spectrometry, 
the sample is counted in a setup with two HPGe detectors  
that can be set to trigger on coincident events (i.e. logical AND). 
The setup may also be set to trigger on either detector (i.e. logical OR) to allow for cross-check with single detector data
and for simultaneous quantification of other nuclides of interest not suited for coincidence counting.
In either counting scheme, the time-dependent observable number of counts in some energy region of interest can be described by the following equation:
\begin{equation}
\label{eq:C}
C(t_i, t_f) = \sum_j A_{0,j} \cdot \tau_j \cdot \varepsilon_j \cdot  (e^{\frac{-t_{i}}{\tau_j}}-e^{\frac{-t_{f}}{\tau_j}} ) 
\end{equation}
where
$C(t_i,t_f)$ is the number of counts received in the energy region of interest 
in the time defined by $t_i$ and $t_f$, which are, respectively, the starting and ending time of counting (where the end of \removed{the} irradiation is defined as time zero), and
$\varepsilon_j$ is the generalized detection efficiency 
which includes the tabulated intensity of the emitted gamma ray, the geometrical acceptance of the setup, the HPGe detector full-energy peak efficiency, and the electronics dead time.
The detection efficiency for nuclide $j$, $\varepsilon_j$, which depends on the counting method and the energy region of interest, 
can be calculated by Monte Carlo particle transport simulations.

\subsection{Determination of trace element concentrations}
\label{sec:ana}
In the final step of the analysis, the activities of the activation products, and hence the concentrations of the trace elements, are deduced from the counting data.
In single gamma spectrometry, 
the energy spectrum near the region of interest is fitted with the sum of a Gaussian function and a quadratic background:
\begin{equation}
f(E) = \frac{C'}{\sigma\sqrt{2 \pi}} \cdot e^{-\frac{(E-\mu)^2}{2\sigma^2}} + a_0 + a_1 \cdot (E - \mu) + a_2 \cdot (E-\mu)^2
\end{equation}
where
$C'$ is the measured counts due to the nuclide of interest,
$E$ is the detected energy, 
$\mu$ and $\sigma$ are the centroid and the spread of the Gaussian function, and
$a_0$, $a_1$, and $a_2$ are nuisance parameters describing the background the peak is superimposed onto. 
In the fit, $\mu$ and $\sigma$ are fixed, respectively, to the known energy of the gamma line and the known energy resolution at that energy. 
As for $\gamma$-$\gamma$ coincidence spectrometry,
the two-dimensional energy spectrum is fitted with the sum of a two-dimensional Gaussian function and a linear background as follows:
\begin{equation}
g(E_A, E_B) = \frac{C'}{2 \pi \sigma_A\sigma_B} \cdot e^{-\frac{(E_A-\mu_A)^2}{2\sigma_A^2} -\frac{(E_B-\mu_B)^2}{2\sigma_B^2}} + b_0 + b_1 \cdot (E_A -\mu_A) + b_2 \cdot (E_B - \mu_B)
\end{equation}
where 
$E_A$ and $E_B$ are the energies detected by detector A and detector B respectively,
$\mu_A$ and $\sigma_A$ ($\mu_B$ and $\sigma_B$), are the centroid and the spread of the energy spectrum detected by detector A (B), 
$b_0$, $b_1$, and $b_2$ are nuisance parameters of the background. 
A detailed bias analysis, utilizing MC truth data, showed that a linear background model suffices for coincidence counting but not singles counting.
Similar to the one-dimensional fit, $\mu_A$, $\mu_B$, $\sigma_A$ and $\sigma_B$ are fixed to their respective known values.
\added{Note that this functional form implies that $E_A$ and $E_B$ are uncorrelated.
This was confirmed to be true using the simulation results described in the next Section.}
In both counting methods, 
the activity of the activation product of interest, $A_0'$, is deduced from the fitted $C'$ using the following equation:
\begin{equation}
\label{eq:anaa0}
A_0' = \frac{C'}{\tau \cdot \varepsilon \cdot (e^{\frac{-t_{i}}{\tau}}-e^{\frac{-t_{f}}{\tau}} ) }
\end{equation}
where $\tau$ is the mean lifetime of the nuclide of interest and 
$\varepsilon$ is the generalized detection efficiency for the nuclide of interest, similar to $\varepsilon_j$ defined above.
In case of multiple regions of interest, the calculated $A_0'$ are averaged, weighted by the reciprocal of the square of their uncertainties, 
to provide a more accurate estimate of the activity.

In contrast to the time-integrated spectral analysis described here, routine activation at UA usually involves time-differential peak fitting~\cite{EXO200:2017}. This time differential fitting, determining a half-life in addition to the peak energy, greatly enhances the reliability of the radio-nuclide assignment. Since in this Monte Carlo study the nuclide mix underlying the data is known beforehand, this step was removed to reduce computational complexity. It is assumed that the analysis improvement factor due to the observation of $\gamma$-$\gamma$ coincidences is independent of this additional observable that could be constructed for both methods being compared.


\section{Comparison between the two counting schemes}

Toy data was generated with a modified version of GeSim, a GEANT4 model of the HPGe detectors at UA ~\cite{Tsang:2019},
based on a measurement of a sapphire sample, supplied by Saint-Gobain North America, performed in 2017. 
The side activities determined in this measurement defined the background composition for this study. The sensitivity estimate, thus, has been made for a realistic high purity material.
The data was then analyzed with both methods described in the previous section.

Comparison to the actual 2017 singles counting-based analysis verified  
the correctness of the data analysis approach.
The methods' sensitivity and discovery potential, serving as FOMs for the detection limit, were calculated using these analysis results \added{as input}.
The calculation of the FOMs utilizes repeated analyses of multiple random data sets, treating the frequency of occurrence as equivalent to probability.

\subsection{Generation of toy datasets}

\begin{figure}
\centering
\includegraphics[width=\textwidth]{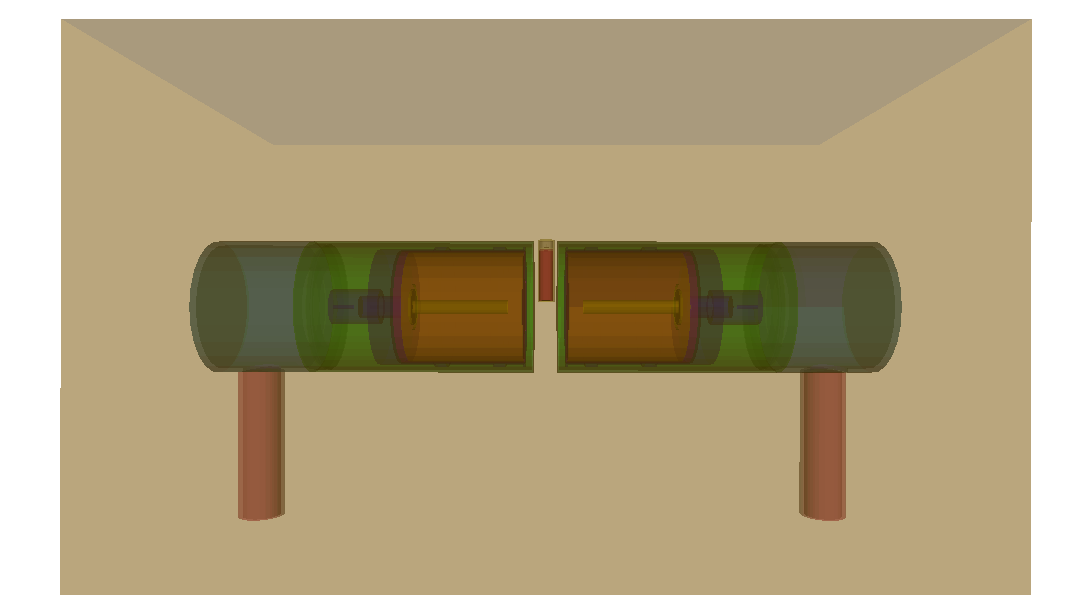} \\
\caption{GEANT4 rendering of the $\gamma$-$\gamma$ coincidence setup investigated in this work. The vial \removed{depicted in the drawing is} located between the two HPGe detectors \removed{to indicate the counting position of}\added{is} the activated sample.} 
\label{fig:2HPGes}
\end{figure}

The toy datasets were generated in two steps: 
first, high-statistics simulations were performed to generate probability density functions (PDF) of the single and coincidence spectra,
then toy datasets were drawn repeatedly from the generated PDFs according to Poisson statistics.
The PDFs were generated with a version of GeSim (GEANT4 10.4.p02), 
which has been modified to contain two HPGe detectors (denoted by Ge-A and Ge-B) 
sitting endcap-to-endcap with a small gap between them for sample placement.
A rendering of the setup is shown in Figure \ref{fig:2HPGes}.
The simulation code was configured to trigger on either detector (i.e. logical OR).
Events where Ge-A was triggered would be extracted as single $\gamma$ data; while 
events where both detectors were triggered (logical AND) would be extracted as $\gamma$-$\gamma$ coincidence data.
In this step, a Gaussian smearing was applied to the energy of the extracted events to mimic the energy resolution of the detectors.
The energy resolution for both detectors, quantified by the standard deviation $\sigma$ of the Gaussian energy response function, was assumed to follow the expression below:
\begin{equation}
\label{eq:res}
\sigma(E) = 0.639338 + 2.80865 \times 10^{-4} \times E 
\end{equation}
where $E$ and $\sigma(E)$ are both in keV.
The coefficients were taken from a recent energy resolution measurement of one of the HPGe detectors at UA.
Before data generation, a bug in $^{99}$Mo decays \removed{has been}\added{was} found in GEANT4. \footnote{\url{https://bugzilla-geant4.kek.jp/show_bug.cgi?id=2387}}
Specifically, the intensity of the 140 keV $\gamma$ line produced by $^{99\text{m}}$Tc, the decay product of $^{99}$Mo, in the simulation was too low compared to the actual data taken at UA and to tabulated nuclear data \cite{lund}. 
While an official bug fix has not been released as of the time of writing, an entry in the Photon Evaporation data version 5.2 has been identified as the culprit and has been fixed.

\begin{table}
\centering
\begin{tabular}{c|ccc}
Nuclide & Half-life & Activity at $t_0$ [Bq] & Number of decays simulated \\
\hline
$^{24}$Na  & 15 h & $ 2.57\times10^5$ & $9.6 \times 10^{10}$ \\
$^{72}$Ga & 14.1 h & $ 1.1\times10^4$ & $10^{10}$ \\
$^{99}$Mo & 65.9 h & $3.0\times10^3$& $10^{10}$ \\
$^{51}$Cr & 27.7 d & 349 & $1.01 \times 10^{10}$ \\
$^{42}$K & 12.32 h & 83 & $10^{9}$ \\
$^{46}$Sc & 83.79 d & 58 & $10^{10}$ \\
$^{181}$Hf & 42.39 d & 7 & $10^{9}$ \\ 
$^{60}$Co & 1925.28 d & 2 & $10^{9}$ \\
$^{239}$Np & 56.5 h & \textit{(Variable)} & $10^9$ \\
\end{tabular}
\caption{Nuclides simulated and their assumed activities at $t_0$, the time of the end of irradiation. 
These were taken from a measurement of a Saint-Gobain G3 sapphire sample provided by the nEXO collaboration in which no evidence of $^{239}$Np was observed.
The activity of $^{239}$Np was set \removed{at}\added{to} various values to study the sensitivity of the analysis methods. 
The last column shows the number of decays simulated for each nuclide.
For $^{51}$Cr, $10^9$ emissions of 320.0824 keV $\gamma$ were simulated, equivalent to about $1.01 \times 10^{10}$ decays given the 9.91\% intensity.
}
\label{tab:simdnuc}
\end{table}

Table \ref{tab:simdnuc} shows the nuclides simulated and their activities as taken from the 2017 single-$\gamma$ NAA measurement of the nEXO sapphire sample mentioned above \cite{sapphirenaa}.
In that measurement, \added{2.5477 g of Saint-Gobain G3 sapphire were activated for 82.81 h at the MIT research reactor, running at a power of 5.70 MW, using their 1PH1 sample insertion facility.
The neutron fluxes were determined using NIST standard fly ash and TiN monitor samples, which were activated for 5.18 minutes immediately following the sapphire irradiation.
The measured neutron fluxes were:
\begin{itemize}
\item $\phi_{th} = (7.11 \pm 0.13) \times 10^{12}~cm^{-2}s^{-1}$
\item $\phi_{epi} = (4.00 \pm 0.79) \times 10^{11}~cm^{-2}s^{-1}$
\item $\phi_f = (8.50 \pm 0.15) \times 10^9~cm^{-2}s^{-1}$
\end{itemize}
A total of about 38 hours elapsed between the end of irradiation and the start of counting at UA.}
No evidence of $^{239}$Np, the activation product of $^{238}$U, was observed. 
Instead, a 90\% C.L. upper limit of 1.1 Bq has been set for the $^{239}$Np activity.

\begin{figure}
  \begin{subfigure}{0.5\textwidth}
    \includegraphics[width=\textwidth]{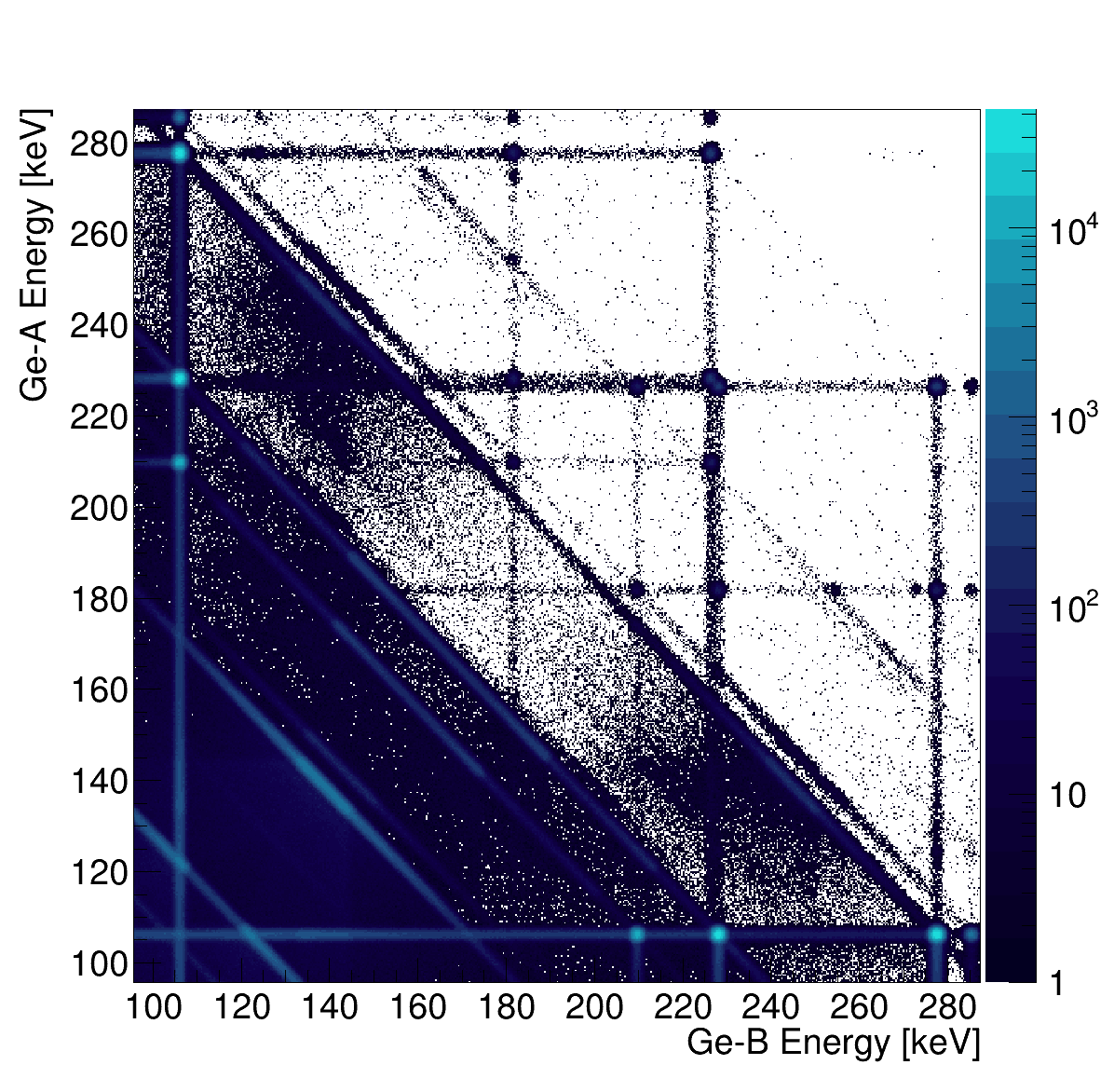}
    \caption{$^{239}$Np}
  \end{subfigure}
  \begin{subfigure}{0.5\textwidth}
    \includegraphics[width=\textwidth]{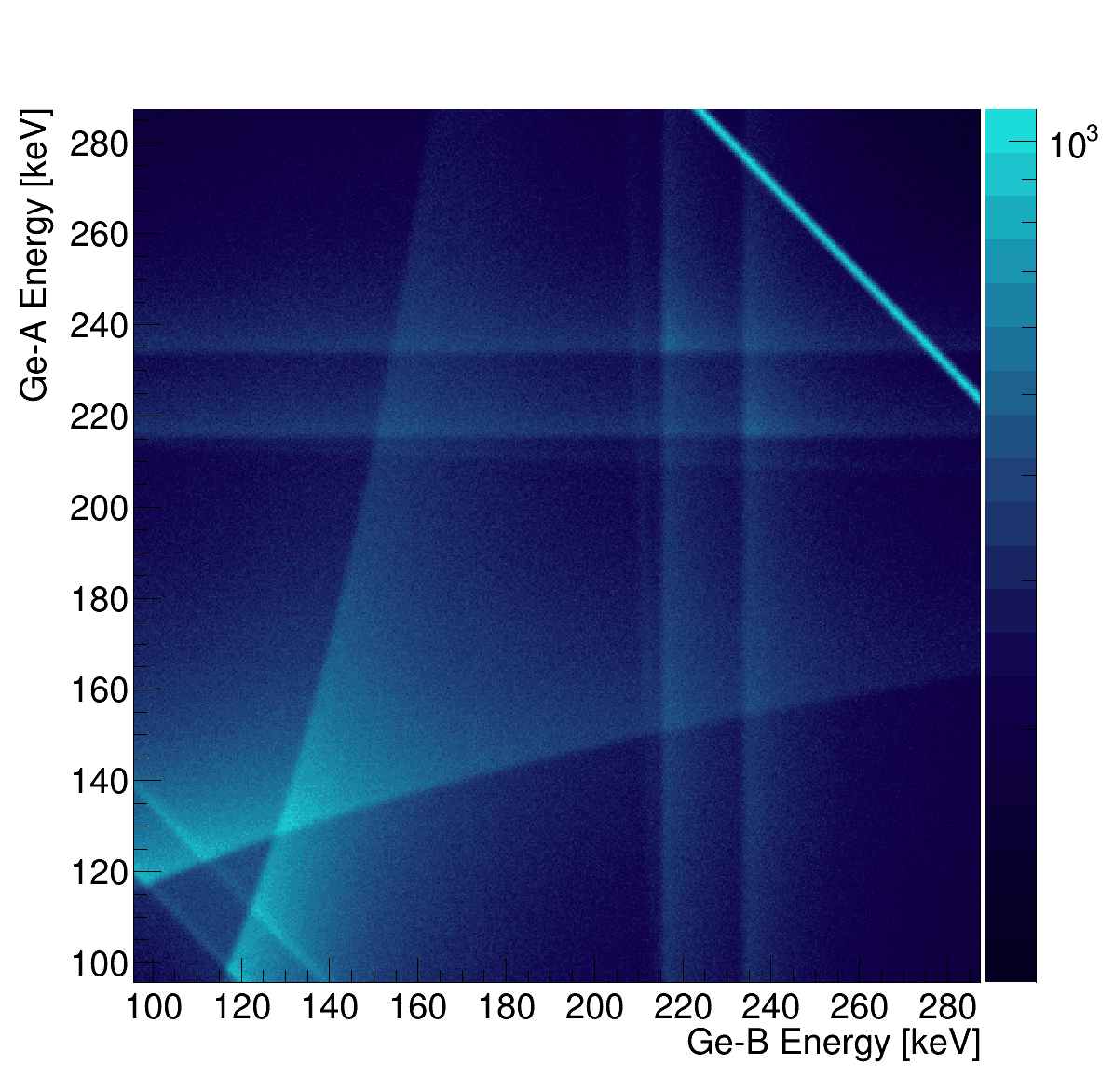} \\
    \caption{$^{24}$Na}
  \end{subfigure}
  \begin{subfigure}{0.5\textwidth}
    \includegraphics[width=\textwidth]{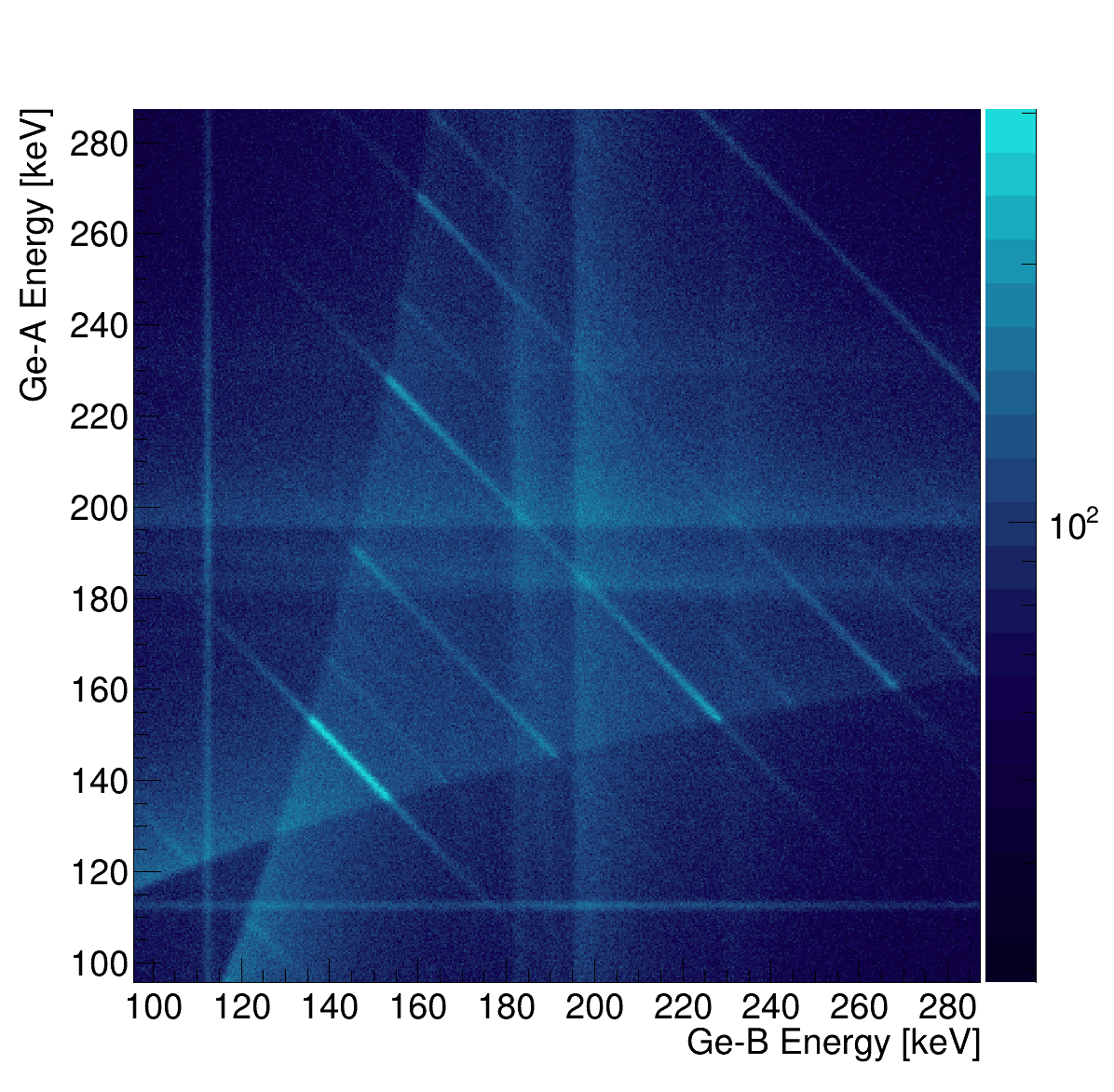} \\
    \caption{$^{72}$Ga}
  \end{subfigure}
  \begin{subfigure}{0.5\textwidth}
    \includegraphics[width=\textwidth]{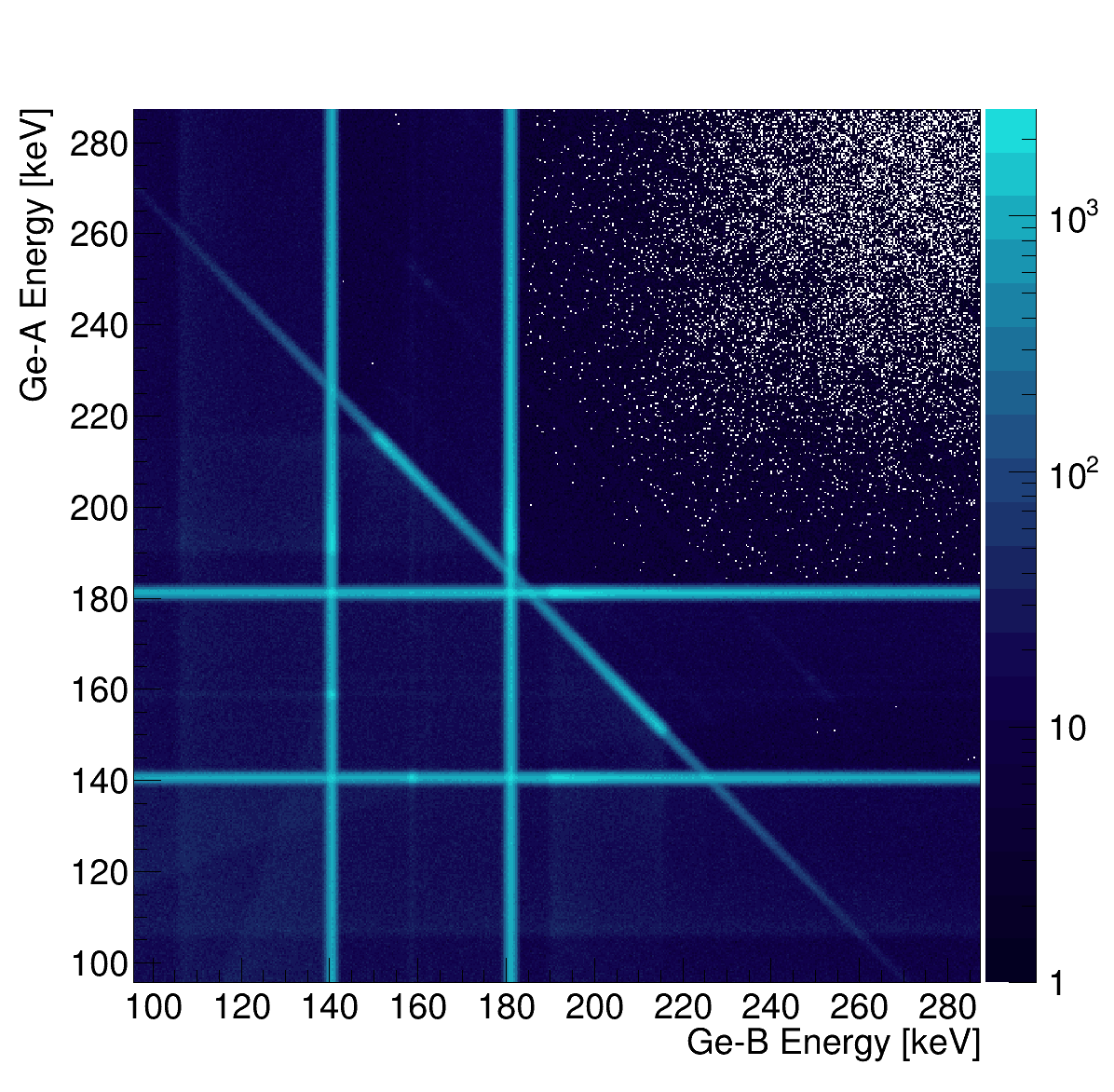} \\
    \caption{$^{99}$Mo}
  \end{subfigure}
  \caption{Raw simulated $\gamma$-$\gamma$ coincidence spectra of $^{239}$Np and the background nuclides $^{24}$Na, $^{72}$Ga, \added{and} $^{99}$Mo.
In the $^{239}$Np spectrum, the cascades of 
the $\frac{7}{2}^- \rightarrow \frac{5}{2}^+$ transition (E = 106.1 keV, I = 25.34\%) followed by either 
the $\frac{5}{2}^+ \rightarrow \frac{5}{2}^+$ (E = 228.2 keV, I = 10.73\%) or 
the $\frac{5}{2}^+ \rightarrow \frac{3}{2}^+$ transition (E = 277.6 keV, I = 14.51\%) are clearly visible as 
four prominent spots.
Two additional spots are identified as the coincident cascades from the $\frac{7}{2}^- \rightarrow \frac{5}{2}^+$ (E = 106.1 keV, I = 25.34\%) transition to $\frac{5}{2}^+ \rightarrow \frac{7}{2}^+$ transition (E = 208.9 keV, I = 3.36\%).
}
  \label{fig:np2dpdf}
\end{figure}

To ensure that sufficient statistics would be accumulated, 
at least $10^9$ decays were simulated for each nuclide. 
For nuclides with particularly high starting activities and/or relatively long half lives more decays were simulated. This was done to to reduce statistical uncertainty.
The number of simulated decays are listed in Table \ref{tab:simdnuc}.
The raw simulated spectra of $^{239}$Np and the three most important background nuclides are shown in Figure \ref{fig:np2dpdf}. 


\begin{figure}
\centering
\includegraphics[width=\textwidth]{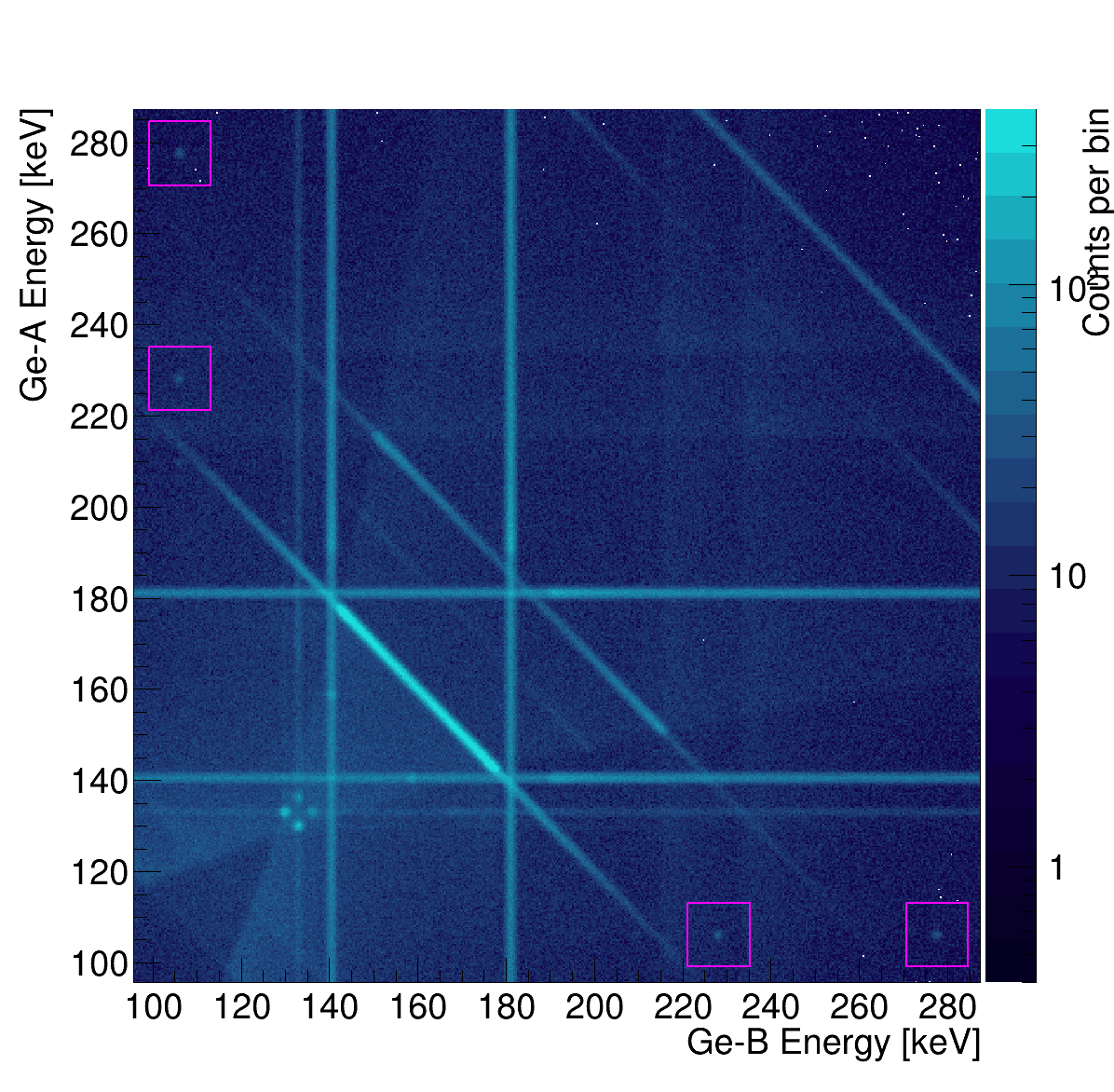} \\
\caption{$\gamma$-$\gamma$ coincidence spectrum of a toy dataset generated with $^{239}$Np activity set at 3 Bq. The four regions of interest are indicated by the magenta boxes.}
\label{fig:data2d}
\end{figure}

\begin{figure}
\centering
\includegraphics[width=\textwidth]{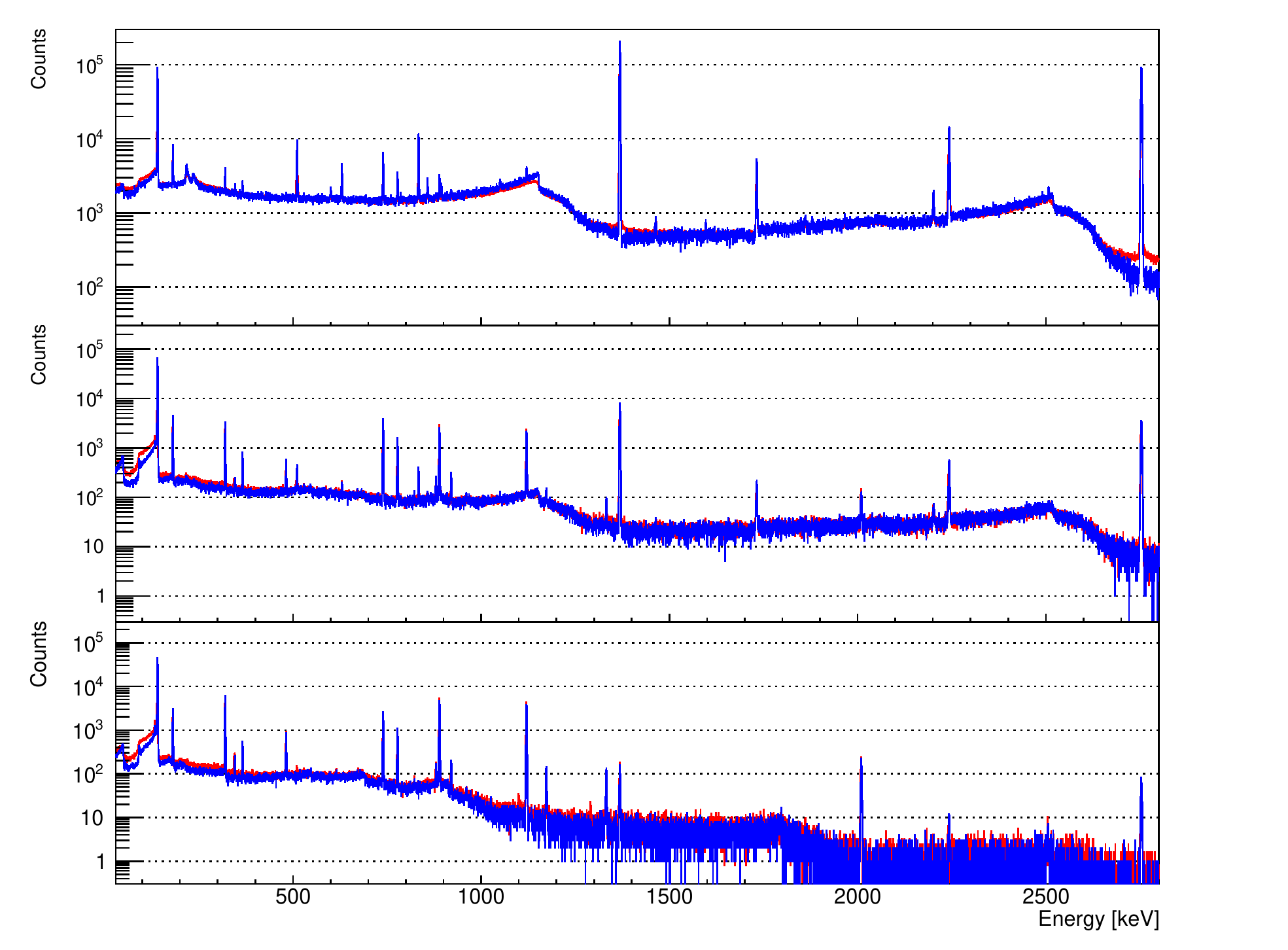}\\
\caption{Single $\gamma$ spectra of a toy dataset (blue) compared with actual data (red) of the sapphire sample taken at three different time intervals. 
From top to bottom: 65.5 to 66.7 hours, 148.1 to 150.1 hours, and 246.8 to 251.8 hours after the end of irradiation. 
\removed{Note that there are two differences between the actual data taking and the simulation: the sample placement and the presence of the second Ge crystal.}\added{For this comparison, two changes in the simulation have been made to match the actual data-taking configuration: the sample vial in the simulation has been moved to the top of the end cap; and the second Ge detector has been removed from the simulation. A small discrepancy between the two spectra can be seen near 100 keV. Its effect on the detection limits has been evaluated and found to be small compared to the uncertainty due to simulation statistics.}
}
\label{fig:datatime}
\end{figure}

Toy datasets were created by randomly generating an appropriate number of events ($C_{trig}$) from the PDFs of each nuclide as described by 
the following equation, 
\begin{equation}
\label{eq:coeff}
C_{trig} = A_0 \cdot \tau \cdot \varepsilon_{trig} \cdot \left(e^{\frac{-t_{i}}{\tau}}-e^{\frac{-t_{f}}{\tau}} \right) 
\end{equation}
in which $t_i$ was set to 35 h, $t_f$ to 371 h, accounting for a cooldown of 35 hours for shipping of the sample from the reactor to UA followed by 14 days of counting.
$\varepsilon_{trig}$ is the probability of triggering the OR trigger, whose value was extracted directly from the simulation results by calculating the ratio of the number of triggered events over the number of simulated decays. 
In the calculation of $\varepsilon_{trig}$, a dead time of 30 $\mu s$ was assumed. 
This value was empirically determined for the existing counting system at UA.
The effect of random coincidences has been evaluated using 200 toy datasets with $^{239}$Np activity set to 3 Bq and random coincidence enabled. The means of the fitted activities for singles and coincidence counting were found to be reduced by $(1.1 \pm 1.2)\%$ and $(0.5 \pm 0.2)\%$, respectively, compared to the toy datasets where random coincidence was disabled. Hence, the effect of random coincidences was considered negligible and excluded from the simulation.
The selected events were then filled into 1-dimensional histograms as single $\gamma$ datasets, and into two-dimensional ones as $\gamma$-$\gamma$ coincidence datasets. 
The bin width of the histograms was 0.348586 keV similar to how the ADC of the actual Ge detector at the University of Alabama was configured.
For better computational performance, the range of the 2D histograms \removed{were}\added{was} limited to $[96.558,288.28]$ keV or $[277,827]$ channels in ADC in both dimensions, 
while for the one-dimensional histograms, the full range of 8192 bins, ranging from 0 to 2855.62 keV\removed{ were}\added{, was} kept.
Shown in Figure \ref{fig:data2d} is an example of a $\gamma$-$\gamma$ coincidence spectrum.
Figure \ref{fig:datatime} shows the single $\gamma$ spectrum at different time intervals compared to actual data taken at UA. 
The agreement between data and Monte Carlo serves to validate the simulation.
$1000$ toy datasets, in which $^{239}$Np activity was set to zero, have been generated. 
$1000$ toy datasets each have been generated for 12 different $^{239}$Np activities ranging from 0.01 Bq to 3 Bq.


\subsection{Analysis of toy datasets}

\begin{figure}
  \begin{subfigure}{0.45\textwidth}
    \includegraphics[width=\textwidth]{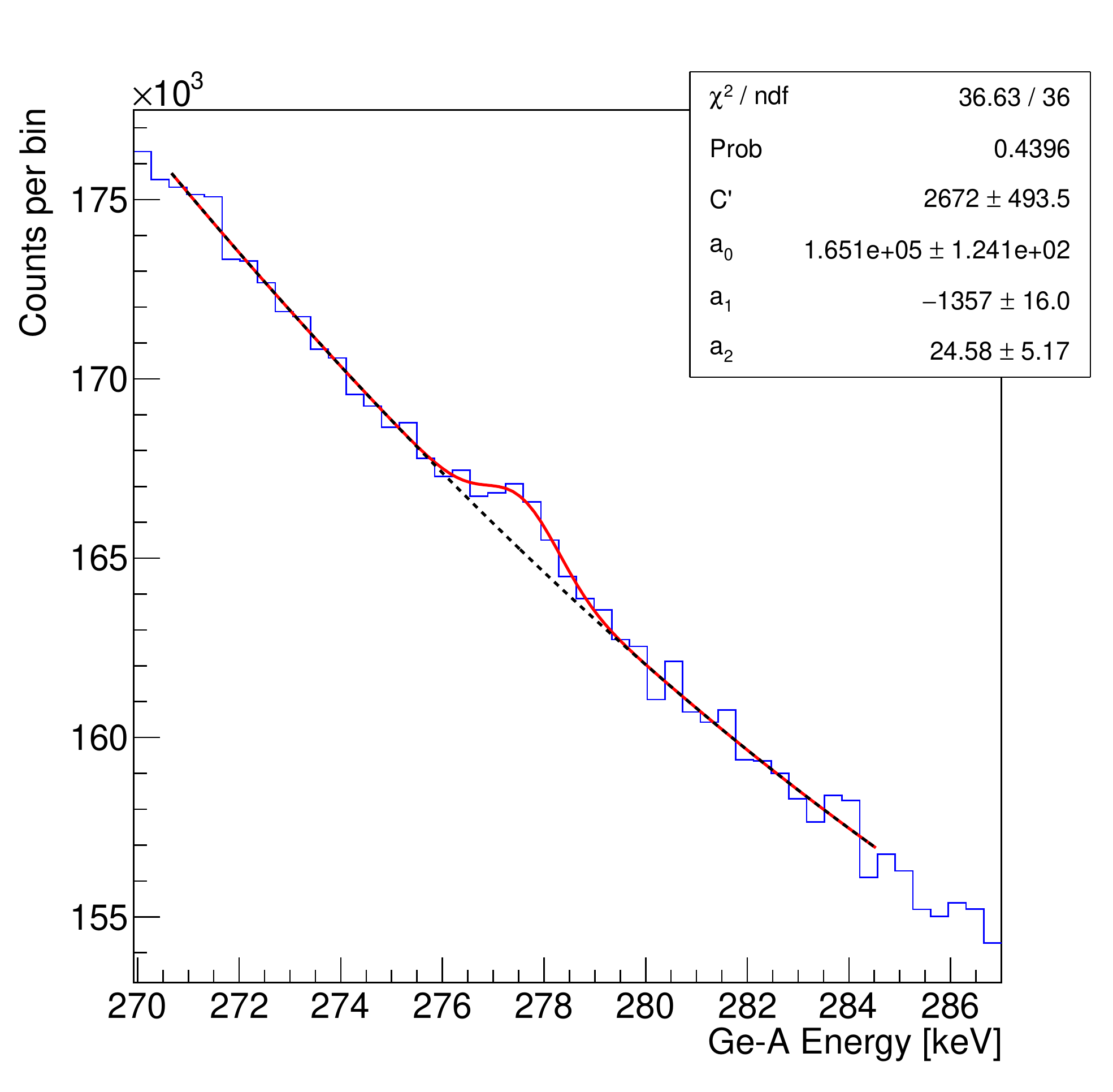} \\
    \caption{Single $\gamma$}
  \end{subfigure}
  \begin{subfigure}{0.55\textwidth}
    \includegraphics[width=\textwidth]{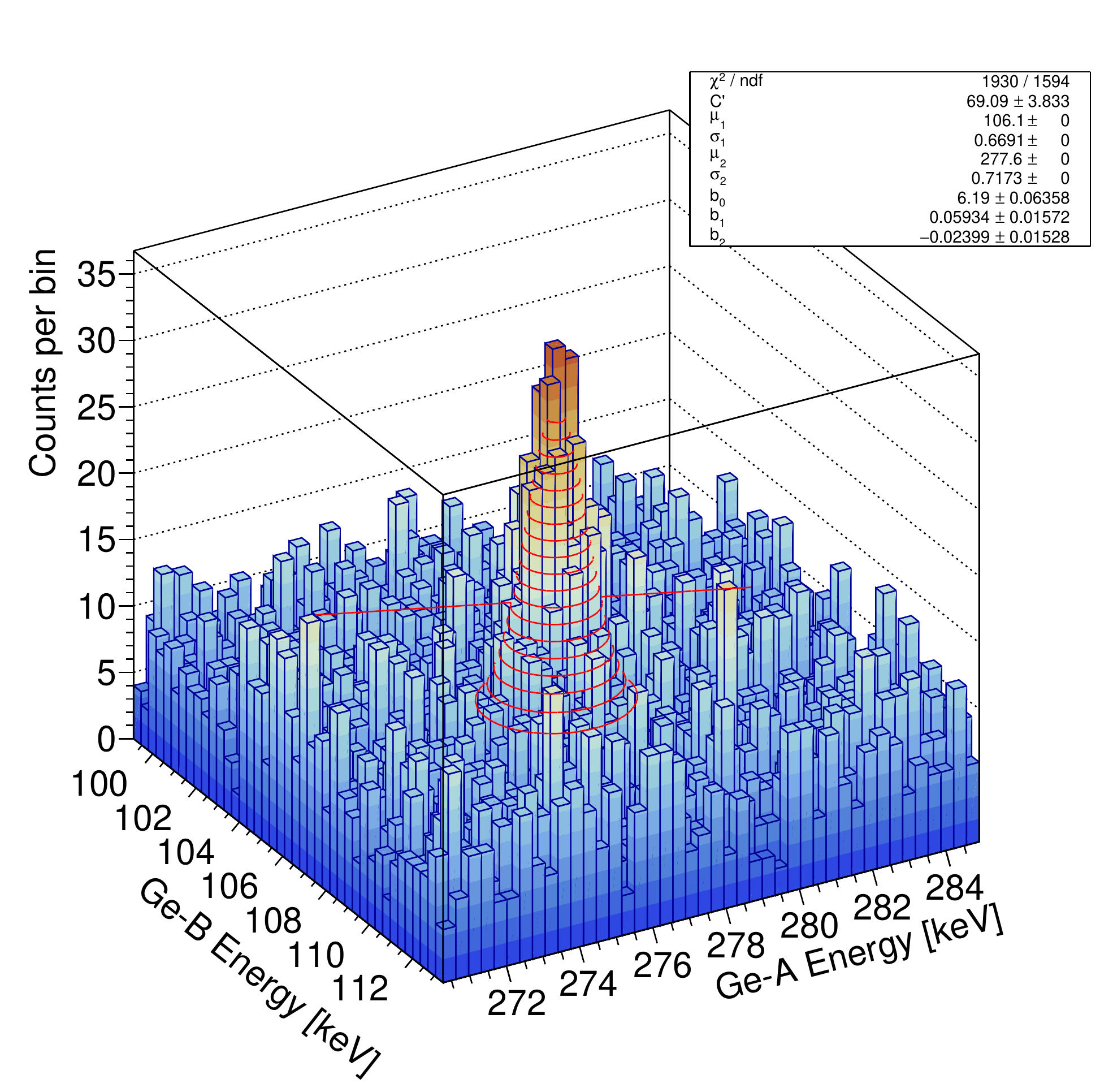} \\
    \caption{$\gamma$-$\gamma$ coincidence}
  \end{subfigure}
\caption{
Examples of a single $\gamma$ peak fit (left) and a $\gamma$-$\gamma$ coincidence peak fit (right) of a toy dataset where the $^{239}$Np was set to 3 Bq.
For the left plot, 
the red curve is the fitted $f(E)$ and the black dotted line is the fitted continuum background $a_0 + a_1 \cdot (E-\mu) + a_2 \cdot (E-\mu)^2$.
The value of $C'$ shown in the inset panel translates to $A_0' = 2.7 \pm 0.5$ Bq using Equation \ref{eq:anaa0}.
For the right plot, 
the red contours show the fitted $g(E)$. 
The value of $C'$ shown in the inset panel translates to $A_0' = 3.0 \pm 0.2$ Bq using the same equation with a different $\varepsilon$.
}
\label{fig:ex1dfit}
\end{figure}

\begin{figure}
  \begin{subfigure}{0.5\textwidth}
    \includegraphics[width=\textwidth]{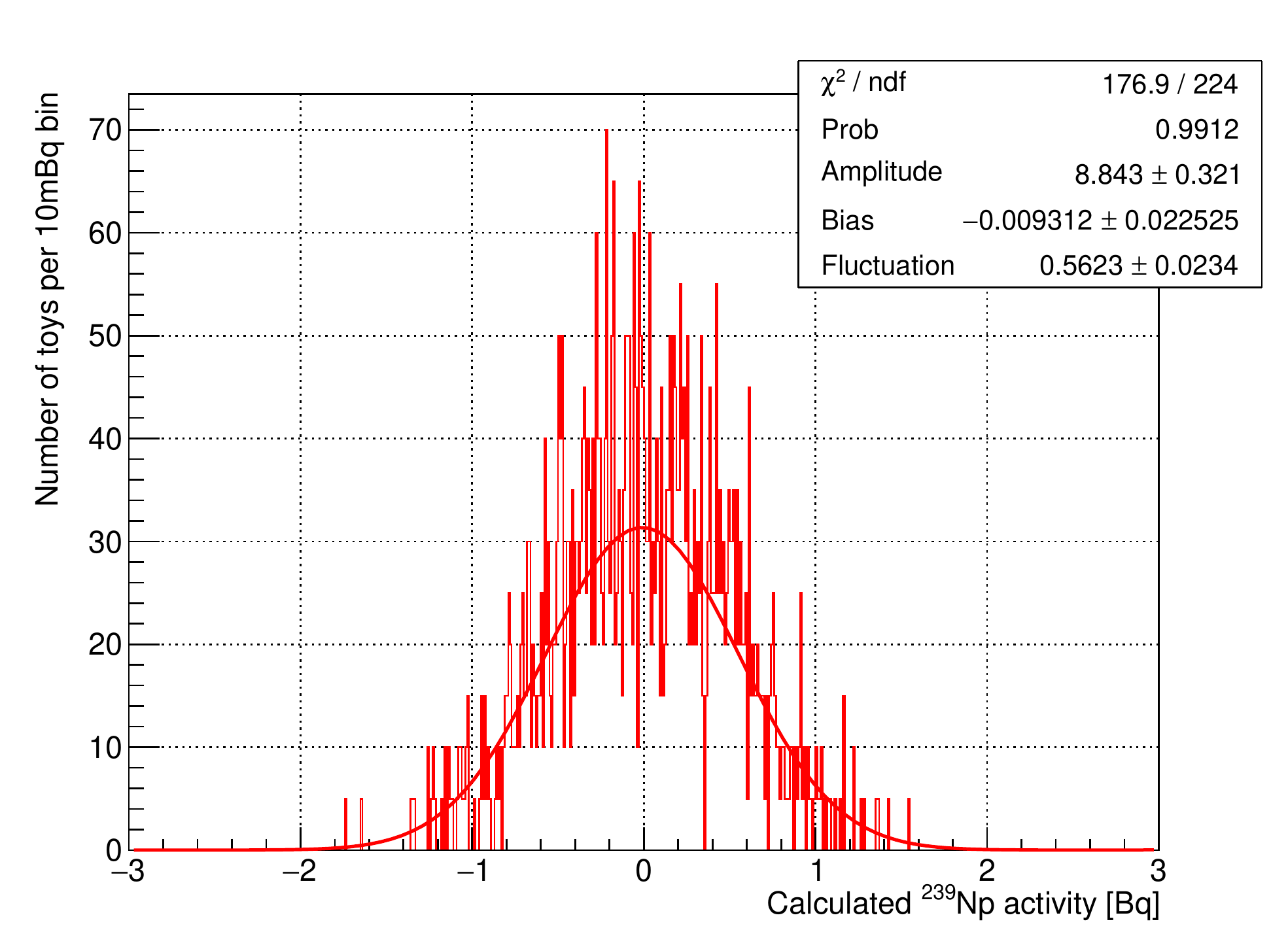} \\
    \caption{Single $\gamma$}
  \end{subfigure}
  \begin{subfigure}{0.5\textwidth}
    \includegraphics[width=\textwidth]{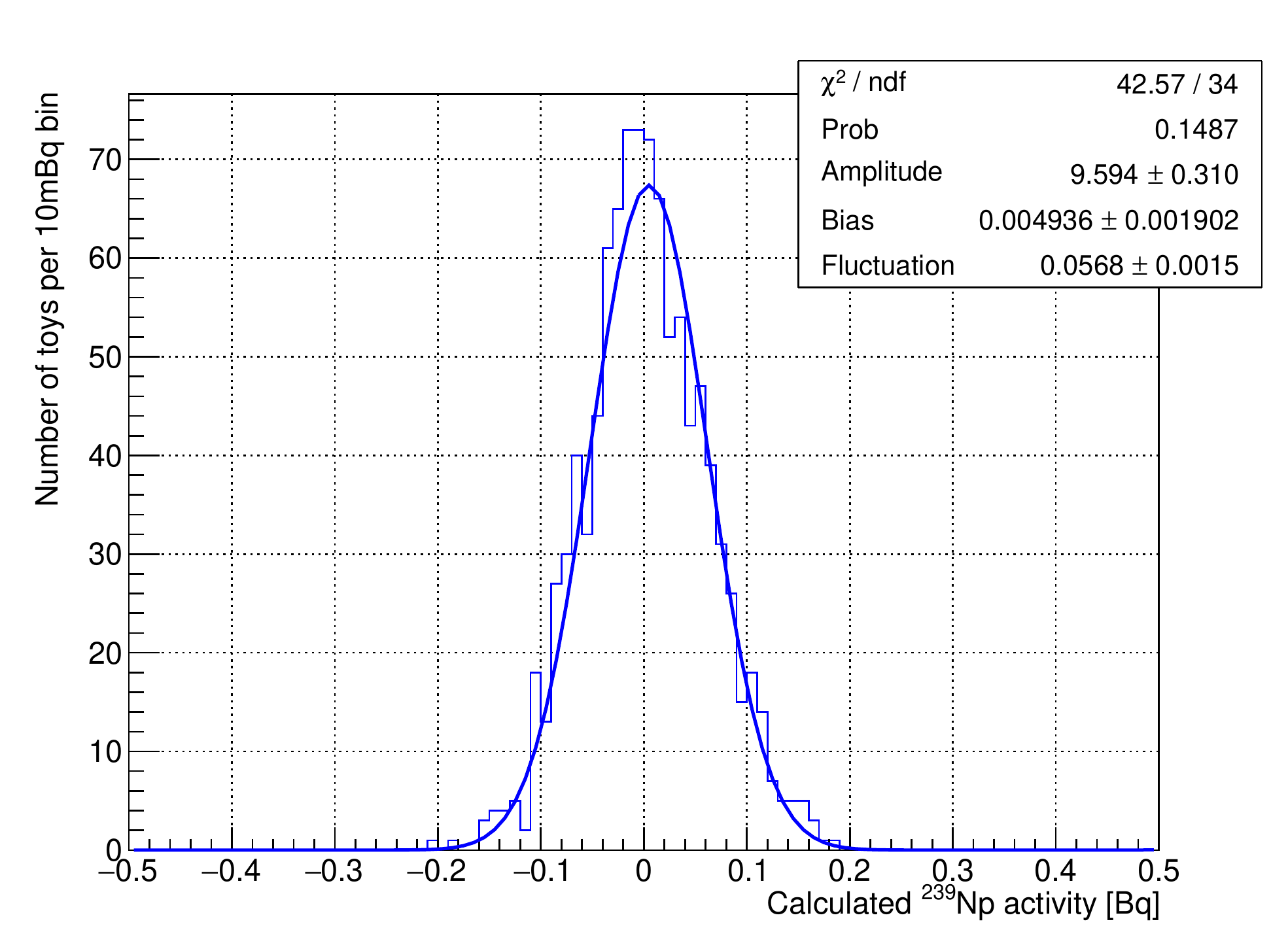} \\
    \caption{$\gamma$-$\gamma$ coincidence}
  \end{subfigure}
  \caption{Systematic biases and statistical uncertainty determined by a Gaussian fit over the fitted $^{239}$Np activities \added{for the case where the true $^{239}$Np was set to zero}.}
  \label{fig:biasfit}
\end{figure}

The single and coincidence spectra of the toy datasets have been analyzed using the methods described in Section \ref{sec:ana}.
For the single $\gamma$ analysis, only the 277.599 keV peak was subjected to the fit.
The fit range was set to $\pm$7 keV
around the 277.599 keV peak energy.
An example fit is shown in Figure \ref{fig:ex1dfit}.
The 228.153 keV peak was not fitted because of its non-Gaussian shape due to a nearby peak at 227.83 keV also from $^{239}$Np.
The 106.123 keV peak was also excluded because it overlaps with a backscatter peak from $^{99}$Mo, making the continuum background near the peak bumpy and hard to model. 
For the $\gamma$-$\gamma$ coincidence analysis, 
the regions within $\pm$7 keV in both dimensions around the following four points:
(106.123, 277.599) keV, (106.123, 228.183) keV, (277.599, 106.123) keV, and (228.183, 106.123) keV, 
were selected as the regions of interest (ROI).
After analyzing all toy datasets, the systematic biases and statistical uncertainties 
of the fitted $A_0'$ were estimated by a Gaussian fit via a frequency distribution of the obtained $A_0'$ values, 
as illustrated in Figure \ref{fig:biasfit} \added{for the case of zero $^{239}$Np activity}.
\added{Similar fits were performed for all other activities and compared to the  Monte Carlo truth.}
In all cases, the uncertainty was found to be dominated by statistics.

\subsection{Detection limit}
\begin{figure}
  \begin{subfigure}{0.5\textwidth}
\includegraphics[width=\textwidth]{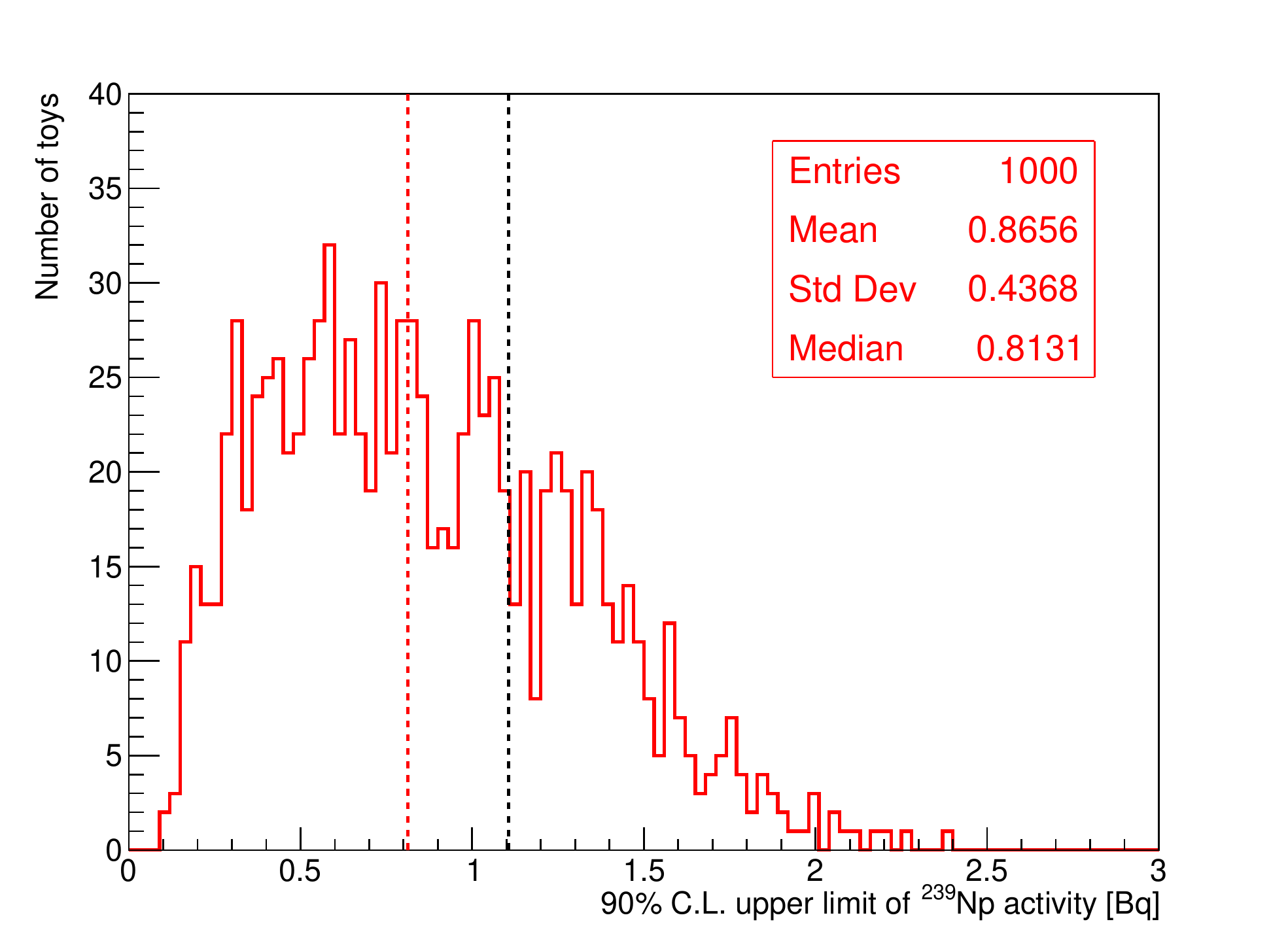}\\
    \caption{Single $\gamma$}
  \end{subfigure}
  \begin{subfigure}{0.5\textwidth}
\includegraphics[width=\textwidth]{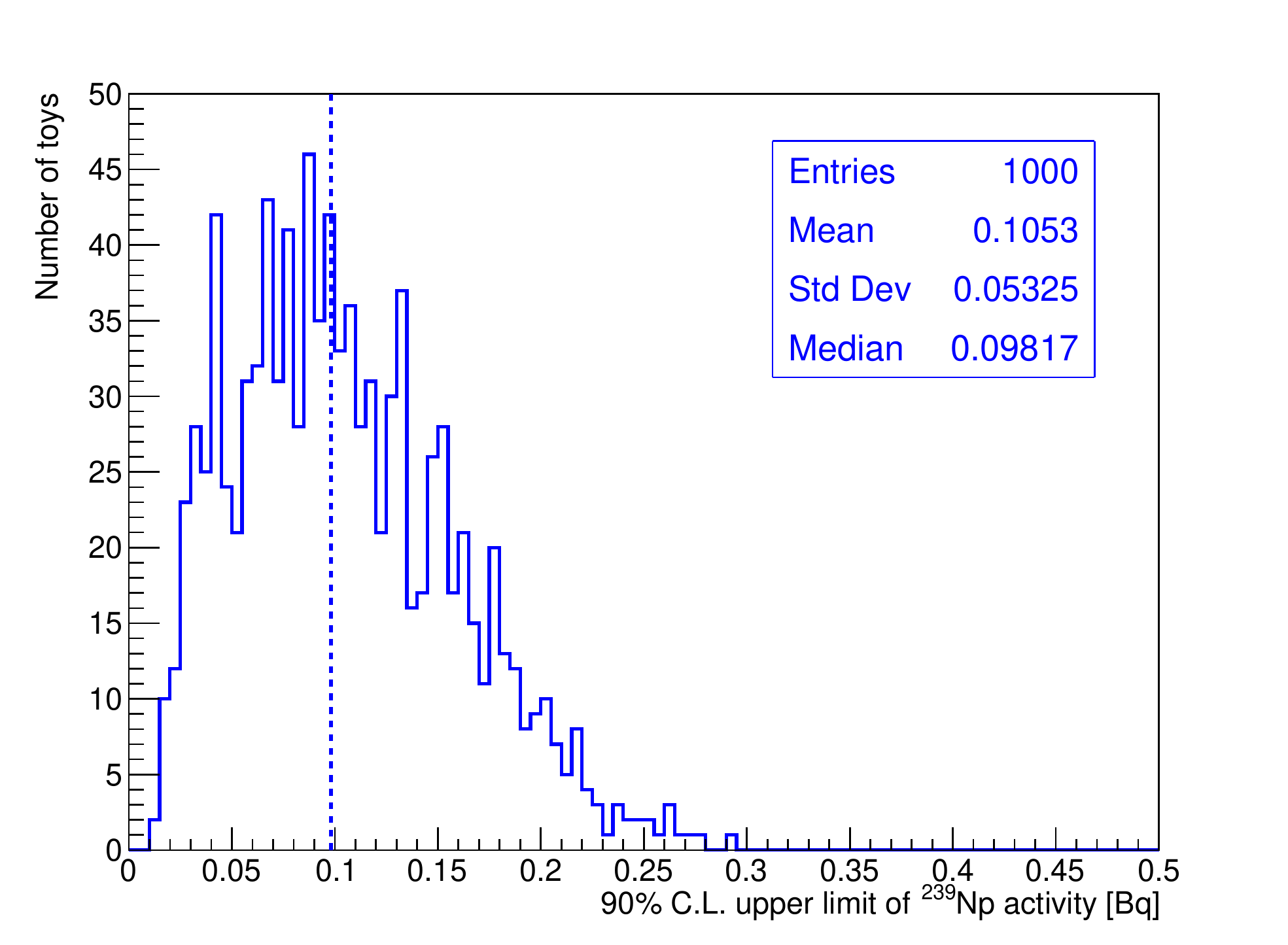}\\
    \caption{$\gamma$-$\gamma$ coincidence}
  \end{subfigure}
\caption{90\% C.L. upper limits set by applying the Feldman and Cousins' method \cite{FC:1998} assuming the null hypothesis for both analysis methods. 
The medians of the distributions for single $\gamma$ (red) and $\gamma$-$\gamma$ coincidence (blue), 
indicated by dotted lines of their corresponding colors in the plot, 
are their respective sensitivities. Also shown in the left plot as a black dotted line is the upper limit of 1.1 Bq set by the actual sapphire measurement made in 2017.}
\label{fig:sens}
\end{figure}

\begin{figure}
\centering
\includegraphics[width=\textwidth]{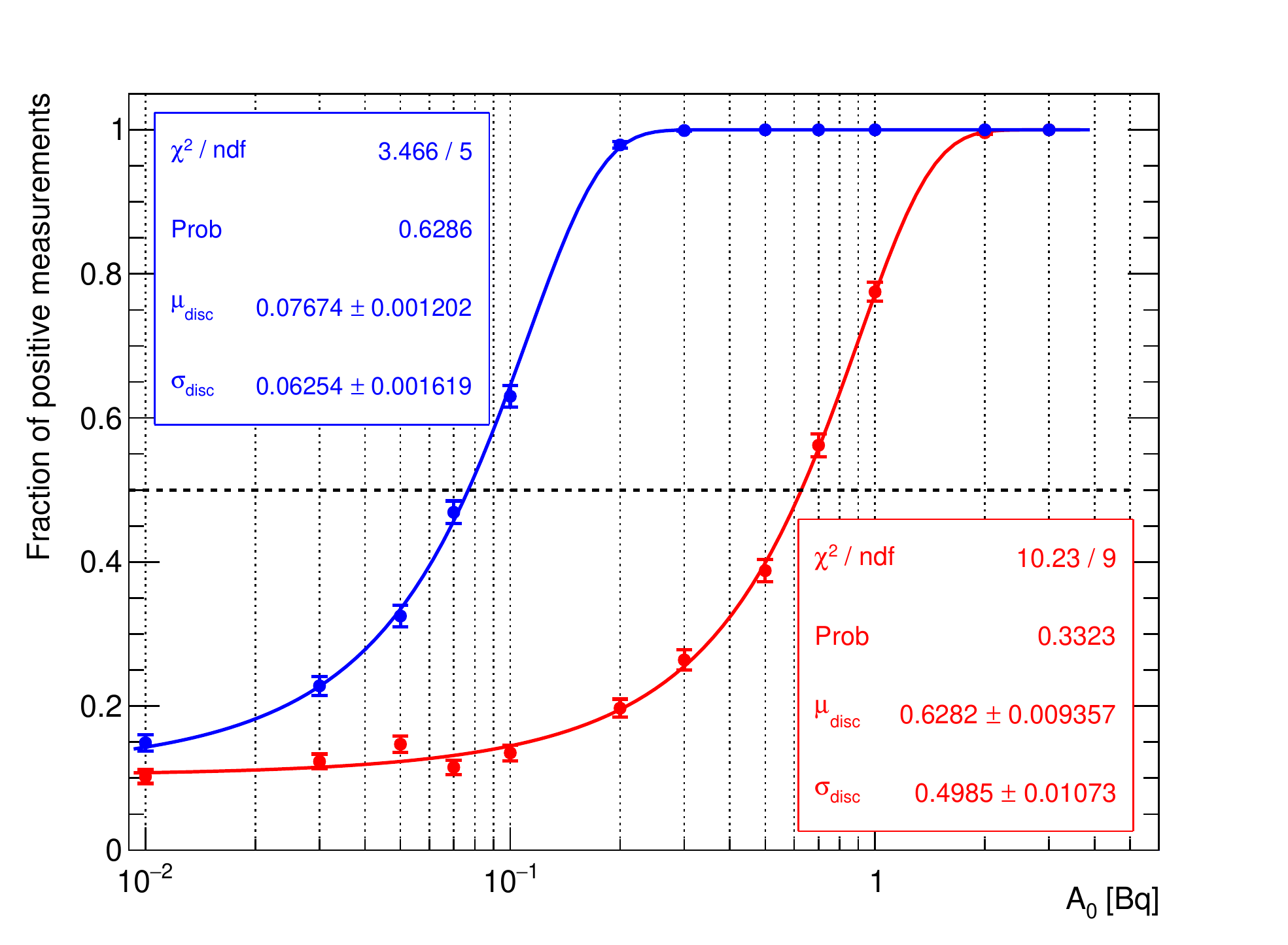}\\
\caption{Fractions of simulated measurements resulting in discovery for the two analysis methods, single $\gamma$ (red) and $\gamma$-$\gamma$ coincidence (blue). 
Their respective results of the Gaussian CDF fits are shown in the insets of the same color.
}
\label{fig:dispot}
\end{figure}

Two figures-of-merit (FOM) have been evaluated as measures of detection limit: sensitivity and discovery potential.
Sensitivity is the median of the 90\% C.L. upper limits that can be set \added{using the Feldman and Cousins' approach \cite{FC:1998}} assuming no $^{239}$Np exists in the sample;
while discovery potential is the $^{239}$Np activity at which the probability of making a 90\% C.L. positive observation of $^{239}$Np is 50\%
when $^{239}$Np of variable but known strength is present in the Monte Carlo data.
Figure \ref{fig:sens} shows the distributions of the upper limits of single and coincidence counting when performing repeated analyses of the null-hypothesis toy datasets. 
Figure \ref{fig:dispot} shows the discovery potentials of the two analysis methods
by scanning the $^{239}$Np activity from 0.01 Bq to 3 Bq.
The $^{239}$Np activity at which the probability of discovery is 50\% was estimated by fitting the points in Figure \ref{fig:dispot} to the Gaussian cumulative distribution function (CDF):
\[
F(x) = \frac{1}{2} \cdot \left[ 1 + \text{erf}\left(\frac{x-\mu_\text{disc}}{\sigma_\text{disc} \cdot \sqrt{2}}\right) \right].
\]
The fitted $\mu_\text{disc}$ quantifies the discovery potential.
Table \ref{tab:sum} summarizes the detection limits of the two methods as evaluated using the two FOMs.
It can be seen that $\gamma$-$\gamma$ coincidence analysis is about 8 times better in both sensitivity and discovery potential, when compared to single $\gamma$ counting.

\begin{table}
\centering
\begin{tabular}{c|cc|c}
 & Single & Coincidence & Improvement factor \\
\hline
Sensitivity [Bq] & 0.81 & 0.098 & 8.3 \\
Discovery Potential [Bq] & 0.63 & 0.077 & 8.2 \\
\end{tabular}
\caption{Summary of detection limits as evaluated by the two FOMs. (See Figures \ref{fig:sens} and \ref{fig:dispot}.)}
\label{tab:sum}
\end{table}

A potentially significant source of systematic uncertainty comes from limited simulation statistics
where repeated toy data draws rely on the accuracy of the underlying PDFs.
To evaluate its magnitude, fits were performed to the raw simulation results, as opposed to drawing events from them as in the generation of toy datasets.
First, the raw PDFs were scaled to $C_{trig}$, as defined in Equation \ref{eq:coeff}, 
and then summed to form a combined PDF. Here the $^{239}$Np activity was set to zero.
\added{The statistical uncertainty of each bin of the raw PDFs was also propagated to the combined PDF.
Thus, the combined PDF represents the simulation truth for the null hypothesis, at the highest possible precision given the available simulation statistics.}
The combined PDF was then fitted to $f(E)$ and $g(E_A,E_B)$. Using the fit results, $A_0'$ was calculated for each case using Equation \ref{eq:anaa0}.
The result was $A_0'= -0.02 \pm 0.08$ Bq for single $\gamma$ and $A_0' = 0.00 \pm 0.01$ Bq for $\gamma$-$\gamma$ coincidence.
\added{Both are consistent with zero, the MC truth value. This means that the simulation is accurate.
The uncertainties on the fitted $A_0'$ show the precision of the simulations.}
Thus, the systematic uncertainties due to simulation statistics are 
$\sigma_\text{sim} = 0.08$ Bq for single $\gamma$ and 
$\sigma_\text{sim} = 0.01$ Bq for $\gamma$-$\gamma$ coincidence.
Their effect on the sensitivity and the discovery potential can be evaluated by 
artificially shifting the fitted $^{239}$Np activities by 
$\pm\sigma_\text{sim}$, then 
calculating the sensitivity and the discovery potential with the shifted activities. 
The results of this evaluation is shown in Table \ref{tab:simbias}.
\removed{The sensitivities for both methods change by about 10\%, while for the discovery potentials, the change is about 15\%.}
\added{The changes in both FOMs range from 10\% to 16\% for both methods.}

\begin{table}
\centering
\begin{tabular}{c|cc|cc}
& \multicolumn{2}{c|}{Sensitivity [Bq]} &  \multicolumn{2}{c}{Discovery potential [Bq]} \\
Bias & Single & Coincidence & Single & Coincidence \\
\hline
$-\sigma_\text{sim}$ & 0.73 & 0.088 & 0.71 & 0.088 \\
0 & 0.81 & 0.098 & 0.63 & 0.077 \\
$+\sigma_\text{sim}$ & 0.90 & 0.11 & 0.55 & 0.065 \\
\end{tabular}
\caption{Systematic bias in sensitivity and  discovery potential due to limited simulation statistics. 
\removed{The sensitivities for both methods change by about 10\%, while for the discovery potentials, the change is about 15\%.}
\added{The changes in both FOMs range from 10\% to 16\% for both methods.}}
\label{tab:simbias}
\end{table}

\section{Conclusion}
\label{sec:conc}
In conclusion, 
the frequentist, Monte Carlo based evaluation of NAA data presented here yields a factor of 8 improvement of coincidence over singles $\gamma$-counting. The evaluation has been guided by data obtained from an activated sapphire sample. For the singles counting case, the spectral distributions and analysis results have been verified by that data, greatly enhancing the reliability of this evaluation.


\section*{Acknowledgements}
\removed{This research was supported in part by the DOE Office of Nuclear
Physics under grant number DE-FG02-01ER41166.
The authors would like to acknowledge the nEXO collaboration for providing us the motivation to conduct this study and providing a sapphire sample.
The authors would also like to thank Tamar Didberidze and Mitchell Hughes for their work on the sapphire measurement that has provided valuable input to the study.}
This research was supported in part by the DOE Office of Nuclear
Physics under grant number DE-FG02-01ER41166. The authors would like to acknowledge the nEXO collaboration for providing us the motivation to conduct this study.
The authors would also like to thank Andrea Pocar at the University of Massachusetts Amherst for the procurement of the sapphire sample, Isaac Arnquist and Sonia Alcantar Anguiano at the Pacific Northwest National Laboratory for the cleaning of the sapphire, and Tamar Didberidze and Mitchell Hughes for their work on the sapphire measurement that has provided valuable input to the study.


\begin{thebibliography}{00}

%
%
%
%
%
%

\bibitem{nexo_sens_2018}
J.B. Albert et al. (nEXO Collaboration),
nEXO: Neutrinoless double beta decay search beyond $10^{28}$ year half-life sensitivity,
arXiv: 2106.16243 (2021). \textit{Submitted to Journal of Physics G.}

\bibitem{Kim:1965}
J.I. Kim, A. Speecke, and J. Hoste,
Neutron activation analysis of copper in bismuth by $\gamma\gamma$-coincidence measurement,
Anal. Chim. Acta \textbf{33} (1965) 123-130

\bibitem{Vobecky:1999}
M. Vobeck\'{y} et al.,
Multielement instrumental activation analysis based on gamma–gamma coincidence spectroscopy,
Anal. Chim. Acta \textbf{386} (1999) 181-189

\bibitem{Hatsukawa:2007}
Y. Hatsukawa, et al.,
High-sensitive elemental analysis using multi-parameter coincidence spectrometer: GEMINI-II
J. Radioanal. Nucl. Chem. \textbf{272}  (2007) 273-276

\bibitem{jendl}
K. Shibata, et al., 
JENDL-4.0: A New Library for Nuclear Science and Engineering,
J. Nucl. Sci. Technol. \textbf{48} (2011) 1-30  \url{http://www.ndc.jaea.go.jp/jendl/j40/j40.html}

\bibitem{EXO200:2017}
D.S. Leonard et al. (EXO-200 Collaboration),
Trace radioactive impurities in final construction materials for EXO-200, 
Nucl. Instr. and Methods A \textbf{871} (2017) 169-179

\bibitem{Tsang:2019}
R.H.M. Tsang et al.,
GEANT4 models of HPGe detectors for radioassay, 
Nucl. Instr. and Methods A \textbf{935} (2019) 75-82

\bibitem{lund}
S.Y.F. Chu, L.P. Ekstr\"{o}m and R.B. Firestone,
The Lund/LBNL Nuclear Data Search, Version 2.0, February 1999,
accessed on July 6, 2021.
\url{http://nucleardata.nuclear.lu.se/toi}

\bibitem{sapphirenaa}
T. Didberidze, M. Hughes, and A. Piepke,
Radiopurity Analysis of Silicon and Sapphire for nEXO,
nEXO internal note (2017)

%
\bibitem{FC:1998}
G.J. Feldman and R.D. Cousins,
Unified approach to the classical statistical analysis of small signals,
Phys. Rev. D \textbf{57} (1998) 3873



 





\end{thebibliography}


\end{document}